\documentclass{article}
\usepackage[utf8]{inputenc}

\usepackage{amsmath,amssymb,amsfonts,amsthm}
\usepackage{physics}
\usepackage{dsfont}
\usepackage{graphicx}

\usepackage{tikz}
\usepackage{tikz-cd}
\usepackage{subcaption}

\usepackage{comment}

\usetikzlibrary{arrows,shapes,trees}

\newcommand{\End}{{\rm End}}

\newcommand{\CC}{{\mathbb C}}
\newcommand{\ZZ}{{\mathbb Z}}
\newcommand{\RR}{{\mathbb R}}

\newcommand{\cP}{{\mathcal P}}

\usepackage[left=1in,right=1in,top=1in,bottom=1in]{geometry}

\usepackage{mathtools}
\usepackage{xfrac}

\usepackage{hyperref}

\usepackage{amsmath}
\usepackage{graphicx}
\usepackage[colorinlistoftodos]{todonotes}

\usepackage[sorting = none]{biblatex}

\begin{document}
\title{Braiding Statistics of Vortices in $2+1$d Topological Superconductors from Stacking}

\author{Minyoung You\\California Institute of Technology, Pasadena, CA 91125}

\maketitle

\begin{abstract}
Class D topological superconductors in $2+1$ dimensions are known to have a $\mathbb{Z}_{16}$ classification in the presence of interactions, with $16$ different topological orders underlying the $16$ distinct phases. By applying the fermionic stacking law, which involves anyon condensation, on the effective Hamiltonian describing the topological interaction of vortices in the $p+ip$ superconductor, which generates the $16$ other phases, we recover the braiding coefficients of vortices for all remaining phases as well as the $\mathbb{Z}_{16}$ group law. We also apply this stacking law to the time-reversal invariant Class DIII superconductors (which can themselves be obtained from stacking two Class D superconductors) and recover their $\mathbb{Z}_2$ classification.
\end{abstract}

\section{Introduction}

Topological phases of gapped matter come in two varieties: topological orders and short-range entanglement (SRE). Both kinds of phases cannot be deformed to the trivial system without closing the gap, and both are beyond the Landau symmetry-breaking classification; however, topological orders exhibit topological ground state degeneracy and topological excitations with fractional statistics, whereas SRE phases lack those characteristics and instead exhibit a nontrivial boundary theory. Fractional quantum Hall systems are well-known examples which exhibit topological order \cite{Laughlin2, MooreRead}. SRE phases can be symmetry-protected, such that they become trivial in the absence of symmetry -- in this case they are called symmetry-protected topological (SPT)  phases -- or they can exist in the absence of any symmetry, in which case they are called invertible topological orders.\footnote{Note that we are using the definition of SRE due to Kitaev \cite{Kitaevtalk2}, which merely requires that they are invertible under stacking; some authors use a different definition involving local unitary transformations}

Some systems can be considered both as an SRE and as a topological order depending on the context or the point of view chosen. Conventional superconductors in $2+1$ dimensions are a well-known example. From the fermionic perspective (where we consider the mean-field BCS Hamiltonian), the system is an SRE, as there are no anyonic excitations -- and in fact belongs to the trivial phase as an SRE. On the other hand, if we consider the physical superconductor, taking into account the electromagnetic gauge field, the vortex is no longer a boundary defect but an anyonic excitation of the system, and in fact the superconductor exhibits the same topological order as the toric code \cite{Hansson2004}. This is a bosonized system obtained by gauging fermionic parity. 

This idea can be applied to topological superconductors in $2+1$d to obtain $16$ different topological orders, or topological quantum field theories (TQFT), given by the fusion and braiding rules of the anyonic quasiparticles  \cite{KitaevAnyons}. These correspond to different number of layers of the basic $p+ip$ superconductor mod $16$, given by the Chern number $\nu$. In \cite{KitaevAnyons} the $16$ different TQFTs were computed algebraically from the fusion rules and then matched to the Chern number through bulk-boundary correspondence. We will exploit the dual perspective of fermionic SRE/bosonic topological order to understand these phases from stacking the bulk TQFTs, without having to refer to the boundary. More precisely, we will stack the effective Hamiltonian describing the topological interaction of vortices in the $p+ip$ superconductor and show that we obtain the correct braiding statistics of vortices for each of the $15 $ other phases. We also describe how the $\ZZ_2$ classification of time-reversal invariant Class DIII superconductors comes about from stacking the corresponding bosonic topological orders.

\section[111]{Review of topological superconductors, the $16$-fold way, and anyon condensation }

\subsection{Vortices in topological superconductors}
\label{sec:vortices}

Consider the $p+ip$ superconductor in $2+1$ dimensions \cite{ReadGreen, Bernevig}:
\begin{equation}
    H = {1 \over 2} \sum_p \begin{pmatrix} c^{\dagger}_p & c_{-p} \end{pmatrix} \begin{pmatrix} {p^2 \over 2m} - \mu & 2i \Delta (p_x + i p_y) \\ 
    -2i \Delta^* (p_x -ip_y) & -{p^2 \over 2m} + \mu \end{pmatrix} \begin{pmatrix} c_p \\ c^{\dagger}_{-p} \end{pmatrix}.
\end{equation} 
$\mu > 0 $ gives the trivial phase with Chern number $\nu = 0$, while $\mu <0 $ leads to the topological phase with $\nu = 1$. If we stack $n$ copies of the nontrivial system, we get $\nu = n$. Layers of $p-ip$ superconductors give a negative contribution to the Chern number. These phases together form a group $\ZZ$ under stacking, which is the well-known result for Class D systems in $2+1$ dimensions \cite{Ryuetal}. In this picture, these phases are clearly invertible and hence are SRE phases.

A $p$-wave superconductor in the presence of a vortex (modelled by the winding behavior of the phase of the order parameter, $\Delta = \Delta_0 (r) e^{i \varphi}$ where $\varphi(\theta) = \theta$)  has a zero-energy Majorana solution to the BdG equations. The Majorana zero mode  is exponentially localized to the vortex:


\begin{equation}
    \gamma = \int r dr d\theta ig(r) \left[-e^{i \theta/2} c(r, \theta) + e^{-i \theta/2} c^{\dagger} (r,\theta) \right]
\end{equation}
where $g(r)$ is exponentially localized at $r = 0$. 

If we have two vortices, we obtain two Majorana zero modes $\gamma_1$ and $\gamma_2$, each localized to the respective vortex core. Exchanging these vortices results in \cite{Ivanov}
\begin{align} \gamma_1 \mapsto \gamma_2  \nonumber \\ 
\gamma_2 \mapsto -\gamma_1.
\label{majoranabraid}
\end{align}
This result can be obtained by keeping track of the branch cuts in the order parameter as we exchange the positions of the two vortices. We can also understand the minus sign on one of the Majorana modes as a requirement for the fermionic parity operator $i \gamma_2 \gamma_1$ to be invariant under the exchange.

\subsection[112]{The interacting classification: the $16$-fold way}

In the absence of interactions we have the Chern number invariant $\nu$, which tells us the net number of layers of the $p+ip$ superconductor. In the presence of interactions, this integer classification breaks down to a $\ZZ_{16}$-classification, which is based on the underlying TQFTs, here given in terms of the type of anyon excitation, and their fusion and braiding rules. In particular, they can be distinguished by the the braiding statistics of vortices \cite{ KitaevAnyons, BernevigNeupert}. We shall denote these phases by $\cP_{\nu}$, $\nu = 1, ..., 16$. 

Here we summarize the results from \cite{KitaevAnyons} which will be relevant. $R^{ab}_c$ will denote the braiding coefficient of $a$ and $b$ in  fusion channel $c$, and $M^{ab}_c = (R^{ab}_c)^2$ will denote the phase due to the double exchange of $a$ and $b$ in fusion channel $c$ (also called the monodromy coefficient): when $a$ and $b$ are of different types, only the double exchange yields a topologically invariant phase factor.

When $\nu \in \ZZ_{16}$ is odd, we have the Ising topological order, consisting of three anyons $1, \sigma, \psi$ with the fusion rules
\begin{align}
    \sigma \times \sigma = 1 + \psi \nonumber \\
    \sigma \times \psi = \sigma \nonumber \\
    \psi \times \psi = 1.
    \label{Isingfusion}
\end{align} 
The braiding coefficients are given by: 
\begin{align}
    R_1^{\sigma \sigma} = \theta e^{i \alpha {\pi \over 4}  } \nonumber \\ R_{\psi}^{\sigma \sigma} = \theta e^{-i \alpha {\pi \over 4}}
    \label{isingbraid}
\end{align}
where $\theta := \theta (\nu) =  e^{\pi i \nu /8} $ and $\alpha = (-1)^{(\nu+1)/2}. $

When $\nu$ is even, we have an abelian theory, but the exact fusion rules depend on whether $\nu = 0$ or $2$ mod $4$. If $\nu = 0$ mod $4$, we have the toric code fusion rules, or the $\ZZ_2 \times \ZZ_2$ fusion rules: four anyons $1, e, m, \psi$ with fusion rules 
\begin{align} e \times e = m \times m = \psi \times \psi = 1 \nonumber \\
e \times m = \psi \nonumber \\
e \times \psi = m \nonumber \\
m \times \psi = e.
\end{align} 
The braiding coefficients for vortices ($e$ and $m$) are
\begin{align}
    R_1^{ee} = R_1^{mm} = e^{\pi i \nu /8  } \nonumber \\
    M_{\psi}^{em} = - e^{\pi i \nu /4}.
    \label{Z2Z2braiding}
\end{align}

When $\nu = 2$ mod $4$, we have the $\ZZ_4$ fusion rules: four anyons $1, a, \psi, \bar{a}$ with 
\begin{align}
a \times a = \bar{a} \times \bar{a} = \psi \nonumber \\
a \times \bar{a} = 1 \nonumber \\
a \times \psi = \bar{a} \nonumber \\
\bar{a} \times \psi = a
\label{Z4fusion}
\end{align}
and the braiding coefficients are 
\begin{align}
    R_{\psi}^{aa} = R_{\psi}^{\bar{a} \bar{a}} = e^{\pi i \nu /8}  \nonumber \\
    M_1^{a \bar{a}}
    = e^{-\pi i \nu /4}.
    \label{Z4braiding}
\end{align}

\subsection{Fermionic stacking and anyon condensation}

Ref. \cite{KitaevAnyons} obtained the above results by computing the possible braiding coefficients for each type of fusion rules, and matched each theory to the Chern number by invoking the bulk-boundary correspondence. If we could instead obtain these theories by stacking the basic $\nu = 1$ system, we would have a way to match the Chern number to a given TQFT from a purely bulk picture, and we could also hope for a better understanding between the even and odd phases (e.g. how do we get abelian topological orders by stacking two non-abelian topological orders?) and the structure of the braiding coefficients as a function of $\nu$. In order to carry this out, we first need to review the correct way to stack these systems.

While the TQFT underlying the $p+ip$ superconductor is correctly described by the Ising topological order, naively stacking two Ising TQFTs, $\cP_1 \boxtimes \cP_1$ leads to a theory with $9$ anyons, rather than the correct $\nu = 2$ topological order. These phases are certainly not invertible under this kind of stacking law -- reflecting the fact that, as bosonic systems, they are topological orders rather than SREs. However, as fermionic systems, they are SREs, and under the proper fermionic stacking law they are indeed invertible and form a group. 

For the Ising topological order without extra structure, the fermion is treated as an anyon, in the sense that it corresponds to a nontrivial superselection sector, while in a fermionic theory we should consider the fermion to a local excitation (possible to create or annihilate by a local operator)  \cite{LanKongWen1}. From this perspective, the Ising TQFT or the toric code theory, which contains a fermion, are modular extensions of the trivial fermionic theory consisting only of $1$ and $\psi$ \cite{LanKongWen1}. The correct stacking law is defined in \cite{LanKongWen2}, and in this case reduces to first using the bosonic stacking law and then condensing the $(\psi, \psi)$ particle \cite{BGK, bosoncondensation}.\footnote{In the language of \cite{BGK, OneformGauging}, this corresponds to gauging the $\ZZ_2$ one-form symmetry generated by $(\psi, \psi)$.} \footnote{Ref. \cite{bosoncondensation} also uses condensation of layers of the Ising topological order to obtain the fusion rules and as well as the topological spins of vortices for the $16$-fold way phases. We will however obtain the braiding coefficients in a more direct manner by stacking effective Hamiltonians.} We will denote this fermionic stacking by $\boxtimes_f.$

An anyon which has trivial braiding with itself can be condensed and is called a condensable boson (for a non-abelian anyon, we require that it has trivial self-braiding in at least one of the fusion channels, though we will only have to deal with condensing abelian anyons). After condensation, several things happen: (1) anyons which nontrivial braiding with the condensed boson become confined; (2) anyons related by fusion with the condensed boson are identified; (3) other anyons can split into different anyons \cite{Burnell, BaisSlingerland}.  

Let us illustrate this with an example which will be relevant  \cite{Burnell}. If we (bosonically) stack two theories with Ising fusion rules, we obtain a theory with nine anyons: $(1,1), (1,\sigma), (1, \psi), (\sigma,1), (\sigma, \sigma), (\sigma, \psi), (\psi, 1)$. We can condense $(\psi, \psi)$, which is a boson. Then,
\begin{align}
    (1, \sigma) \sim (1, \sigma) \times (\psi, \psi) = (\psi, \sigma) \nonumber \\
    (\sigma, 1) \sim (\sigma, 1) \times (\psi, \psi) = (\sigma, \psi) \nonumber \\
   (\psi, 1) \sim (\psi, 1) \times (\psi, \psi) = (1, \psi) 
\end{align}
and $(1, \sigma)$ and $(\sigma, 1)$ are confined. Indeed, physically there should be only one gauge field, whose flux through the system creates vortices on both layers. $(\sigma, 1)$ and $(1, \sigma)$ correspond to vortices which independently live on a single layer, and are forbidden in a physical superconductor.

We are left only with $(1,1), (\sigma, \sigma)$, and $(1, \psi)$.  $(1,1)$ clearly takes the role of the vacuum, which we denote by $\tilde{1}$, and $(1, \psi)$ is a fermion, which we denote by $\tilde{\psi}.$

Note that 
\begin{equation}
(\sigma, \sigma) \times (\sigma, \sigma) = (1,1) + (1, \psi) + (\psi, 1) + (\psi, \psi) \sim \tilde{1} + \tilde{1} + \tilde{\psi} + \tilde{\psi}. 
\end{equation}
As $(\sigma, \sigma)$ fuses with itself to two copies of the vacuum in the condensed phase, $(\sigma, \sigma)$ cannot be a single type of anyon -- it actually splits into two anyons. 

One possibility is that it splits as $(\sigma, \sigma) \mapsto e + m$ with
\begin{align}
    e \times e = m \times m = \tilde{1} \nonumber \\
    e \times m = \tilde{\psi}.
\end{align}  
It is easily verified that $(e+m) \times (e +m ) = \tilde{1} + \tilde{1} + \tilde{\psi} +\tilde{\psi}.$ Then, we end up with four anyons $\tilde{1}, \tilde{\psi}, e, m$ with $\ZZ_2 \times \ZZ_2$ fusion rules. 

Another possibility is $(\sigma, \sigma) \mapsto a + \bar{a}$ with the fusion rules 
\begin{align}
    a \times \bar{a} = \tilde{1} \nonumber \\
    a \times a = \bar{a} \times \bar{a} = \tilde{\psi}.
\end{align}
This also satisfies the condition that $(a + \bar{a} ) \times (a + \bar{a} )=  \tilde{1} + \tilde{1} + \tilde{\psi} +\tilde{\psi}.$ Then we obtain a theory with four anyons $\tilde{1}, a, \bar{a}, \tilde{\psi} $ with $\ZZ_4$ fusion rules. 

Which kind of theory we end up with depends on the exact braiding coefficients of the Ising theory we are stacking, and can be determined algebraically: see e.g. \cite{bosoncondensation}. We shall see that in the case of the $16$-fold way, stacking gives a simple and concrete way to determine the fusion rules for even $\nu$.

\section[222]{Effective Hamiltonian for vortices of an odd $\nu$ phase}

\subsection{Braiding coefficients and superselection sectors}

Consider two vortices, with the corresponding Majorana zero modes $\gamma_1$ and $\gamma_2$ respectively. These combine into a single set of creation and annihilation operators, 
\begin{align} a = {1 \over 2} (\gamma_1 + i \gamma_2) \nonumber \\ 
a^{\dagger} = {1 \over 2} (\gamma_1 - i \gamma_2) \end{align}
and act on a Hilbert space $\CC^2$ spanned by $|0\rangle$ and $|1 \rangle$, which are respectively unoccupied and occupied with respect to $a, a^{\dagger}$. 

As discussed in Sec. \ref{sec:vortices}, braiding two vortices results in
\begin{align} \gamma_1 \mapsto \gamma_2  \nonumber \\ 
\gamma_2 \mapsto -\gamma_1.
\end{align}
This can also be derived by the following reasoning: the states $|0\rangle$ and $|1 \rangle$ formed from two Majorana modes differ by a fermion. If we do a $2 \pi$ rotation of the whole configuration, it is equivalent to two braids between $v_1 $ and $v_2$, and hence should give us a $R^2$. On the other hand, a fermion acquires a sign under $2 \pi$ rotation, so $|0\rangle \mapsto |0 \rangle $ and $|1 \rangle \mapsto - |1 \rangle$, i.e. $R^2 = \begin{pmatrix} 1 & 0 \\ 0 & -1 \end{pmatrix} = i \gamma_2 \gamma_1$, i.e. it just acts by fermionic parity and hence reverse the sign of each $\gamma$. This is achieved by a single $R$ taking  $\gamma_1 \mapsto \gamma_1$, $\gamma_2 \mapsto -\gamma_2 $ or vice versa, since $\gamma$s have to be real.

The operators $\gamma_1$ and $\gamma_2$ generate $Cl(2) \simeq Mat(2, \CC)$ which acts on the $\CC^2$ spanned by the states $|0\rangle$ and $|1\rangle$. From the fermionic point of view, which considers fermions to be fundamental particles, the two basis states belong to the same superselection sector. From a bosonic point of view, however, they belong to different superselection sectors:  the bosonic operators $1$ and $\gamma_1 \gamma_2$ are both diagonal in this basis, so there is no way to move from one state to another if we employ only bosonic operators. 

Recall that the Ising topological order describes a bosonized picture of the topological superconductor: the anyon $\psi$ corresponds to a nontrivial superselection sector since there is no local bosonic operator which can create it out of the vacuum $1$. Let us denote the states in the two superselection sectors $1$ and $\psi$, which are the two possibilities we can land on when fusing two $\sigma$ particles (which carry Majorana modes), as $|\sigma \sigma ; 1 \rangle$ and $ | \sigma \sigma ; \psi \rangle $. These should correspond to the states $| 0 \rangle$ and $|1 \rangle$, which are even and odd, respectively, under the fermonic parity $i \gamma_2 \gamma_1$. Since we do not \textit{a priori} know which one is odd and which is even,  we write  \cite{KitaevAnyons}
\begin{align}
    i \gamma_2 \gamma_1 | \sigma \sigma ; 1 \rangle = -\alpha | \sigma \sigma ; 1 \rangle \nonumber \\ 
    i \gamma_2 \gamma_1 | \sigma \sigma ; \psi \rangle = +\alpha | \sigma \sigma ; \psi \rangle 
\end{align}
for some $\alpha = \pm 1$.

The operator on $\CC^2$ which accomplishes Eq. \eqref{majoranabraid} by conjugation is 
\begin{equation}
    R = \theta e^{-{\pi \over 4} \gamma_1 \gamma_2} =  \theta e^{- i {\pi \over 4} (i\gamma_2 \gamma_1)}
    \label{braidingoperator}
\end{equation}
where $\theta$ is a phase factor (which can be interpreted as the topological spin of the $\sigma$ anyon \cite{KitaevAnyons}). By noting how $i \gamma_2 \gamma_1$ acts on the states $| \sigma \sigma ; 1 \rangle$ and $| \sigma \sigma ; \psi \rangle$, we see that  $R_1^{\sigma \sigma} = \theta e^{i \alpha {\pi \over 4}  } $ and $R_{\psi}^{\sigma \sigma} = \theta e^{-i \alpha {\pi \over 4}} $. 

If $\nu = 1$ mod $4$, $\alpha = -1$. Since $i\gamma_2 \gamma_1$ is fermionic parity, this means that $(-)^F |\sigma \sigma ; 1 \rangle = + |\sigma \sigma; 1 \rangle$, i.e. the fusion channel $1$ corresponds to the ``unoccupied'' state $|0 \rangle$; similarly, $\psi$ corresponds to the ``occupied'' state $|1 \rangle$. 

On the other hand, if $\nu = 3$ mod $4$, $\alpha = +1$, and we have $i \gamma_2 \gamma_1 | \sigma \sigma ; 1 \rangle =  - |\sigma \sigma; 1\rangle$, etc. The fusion channel $1$ corresponds to $|1 \rangle$ and $\psi$ to $|0 \rangle$.  

Note that $R$ as an operator acting on $\CC^2$ is fixed to be of the form $\theta e^{-{\pi \over 4} \gamma_1 \gamma_2}$; the difference between $\cP_{4n + 1} $ and $\cP_{4n + 3}$ is in how one interprets the fusion channels in terms of the fermionic states, and this will be important for stacking.

\subsection{The Effective Hamiltonian for two vortices}

Consider an odd $\nu$ system. A Hamiltonian describing the interaction of two $\sigma$ vortices can be written in the form:
\begin{equation}
    H = {1 \over 2m} \left( (\vec{p}_1 - \vec{A}_1)^2  + (\vec{p}_2 - \vec{A}_2)^2\right) + V^{(1)} (|\vec{r}_1 - \vec{r}_{2}|)  + i\gamma_2 \gamma_1 V^{(2)}(|\vec{r}_1 - \vec{r}_{2}| ). 
   \label{EH1}
\end{equation}
This Hamiltonian acts on the Hilbert space $L^2(\RR^2) \otimes L^2(\RR^2) \otimes \CC^2$, where $\CC^2$ is spanned by the occupied and unoccupied fermionic states. 

The kinetic term is determined by the braiding rules. We write
\begin{equation}
    \vec{A}_1 = \frac{F(\nu)}{|\vec{r}_1 - \vec{r}_2|^2} \left( -(y_1 - y_2), x_1 - x_2 \right)
\end{equation}  as the gauge field felt by the vortex $1$ due to vortex $2$. $F(\nu) $ is a $\nu$-dependent factor valued in $\End (\CC^2)$; it takes the form 
\begin{equation}
    F(\nu) =  -{i \over 4}  \gamma_2 \gamma_1 + {\nu \over 8}.
\end{equation}
The expression for $\vec{A}_2$ is similar.

$V^{(1)}$ is some potential term which does not act on the internal $\CC^2$, and $V^{(2)}$ is some Hermitian potential ($i \gamma_2 \gamma_1$ is itself Hermitian) which splits the energies of the two states in $\CC^2$. In \cite{splitting} the splitting energy is calculated to be $V(R) \approx - 2{\Delta_0 \over \pi^{3 \over 2}} \frac{\cos p_F R +{\pi \over 4}}{\sqrt{ p_F R}} e^{-R/\xi}$ for large separation $R \gg \xi$ where $\xi$ is the superconducting coherence length and $\Delta_0$ is the mean-field value of the superconducting order parameter $\Delta.$ This can affect the braiding phases by some non-universal factor, but we will assume that the vortices are far enough apart that we can ignore the potential terms, and focus on the universal properties of their braiding.

In terms of complex coordinates $z = x + i y$, $x = {z + \bar{z} \over 2}$, $y = {z - \bar{z} \over 2i}$ and 

\begin{equation}
    \vec{A}_1 \cdot d\vec{r}_1 = F(\nu) \left( {1 \over 2i} \frac{dz_1}{z_1 - z_{2}} - {1 \over 2i} \frac{d\bar{z}_1 }{\bar{z}_1 - \bar{z}_2}\right).
\end{equation}   
This Hamiltonian is constructed to give the braiding coefficients Eq. \eqref{isingbraid}. If vortex $1$ encircles vortex $2$, which corresponds to a double-braiding, the wavefunction changes by 
\begin{equation}
    e^{i \oint \vec{A}_1 \cdot d\vec{r}_1}  = \exp{2 \pi i F(\nu) } = \exp{ {\pi  \over 2}  \gamma_2 \gamma_1 }e^{i \pi \nu \over 4}
\end{equation}
which produces the correct double-braiding coefficients, i.e. the square of Eq. \eqref{braidingoperator}. Even though we have given the general form of the effective gauge fields and effective Hamiltonian which work for any odd $\nu$, we will see that we can obtain each of them by starting with the $\nu = 1$ system and repeatedly staking layers of it.

When $\nu$ is even, we have multiple types of vortices and hence the effective Hamiltonian describing the interaction of vortices depends on the specific types of vortices we consider. The form of the Hamiltonians for even $\nu$ will be written down when we discuss stacking; see Sec. \ref{Ham2} for a concrete example for $\nu = 2$.

\section{Stacking: even from odd-odd}

\subsection[41]{Stacking two $\nu =1$ systems}
\label{nu11}

Now that we have written down the effective Hamiltonians for odd $\nu$ (hence, in particular, for $\nu = 1$) we shall verify that we can obtain the braiding statistics of vortices of other phases by stacking the $\nu  =1 $ system. We discuss here in detail the case $\cP_2 = \cP_1 \boxtimes_f \cP_1$; all other cases of stacking two odd systems to get an even system follow the same structure. 

First, take the $ \nu =1 $ system with two vortices, of Eq. \eqref{EH1}:
\begin{equation}
    H = {1 \over 2m} \left( (\vec{p}_1 - \vec{A}_1)^2  + (\vec{p}_2 - \vec{A}_2)^2\right) + V^{(1)} ( |\vec{r}_1 - \vec{r}_{2}|) + i \gamma_2 \gamma_1 V^{(2)} (|\vec{r}_1 - \vec{r}_{2}| ). 
\end{equation}

We stack it with the same system; we write the second layer as 
\begin{equation}
    \bar{H} = {1 \over 2m} \left( (\vec{\bar{p}}_1 - \vec{A}_1)^2  + (\vec{\bar{p}}_2 - \vec{A}_2)^2\right) + V^{(1)} (| \vec{\bar{r}}_1 - \vec{\bar{r}}_2 | ) + i \bar{\gamma}_2 \bar{\gamma}_1 V^{(2)} ( | \vec{\bar{r}}_1 - \vec{\bar{r}}_{2} | ) 
\end{equation}
where the bars simply denote that we have different coordinate and momentum variables, as well as different Majorana operators, from the first layer, even though the two are formally the same. The gauge fields $\vec{A}_i$ on the second layer are written in terms of the barred Majorana operators $\bar{\gamma}_i$ and the barred coordinates $\bar{r}_i$.

Stacking these two systems, we obtain 
\begin{equation}
    H' = H \otimes \mathds{1} +\mathds{1} \otimes \bar{H} 
\end{equation}
acting on $(L^2(\RR^2))^{\otimes4} \otimes \CC^4$. $H'$ depends on four coordinates $z_1, z_2, \bar{z}_1, \bar{z}_2$, which are the positions of the first and second vortex on the two layers. Recall that we need to condense the $(\psi, \psi)$ anyon in the stacked phase in order to get to the resultant fermionic phase. This condensation does three things: confinement, identification, and splitting. 

Confinement occurs for the $(\sigma, 1) \sim  (\sigma, \psi)$ and $(1, \sigma) \sim ( \psi, \sigma)$ anyons. This can be achieved by introducing a potential such as $V \sim e^{|z_i -\bar{z}_i|}$, which forces the position of the vortices on each layer to be the same -- there is no way to move $z_i$ independently of $\bar{z}_i$, so $(\sigma, 1)$ and $(1, \sigma)$ are confined.  After confinement, we obtain:
\begin{equation}
H = {1 \over 2m} \left( (\vec{p}_1 - \vec{A}_1)^2  + (\vec{p}_2 - \vec{A}_2)^2\right) + (\text{potential terms}) \end{equation}
where $\bar{\gamma}_1 $ and $\bar{\gamma}_2$ are the Majorana modes of the second $\nu = 1$ layer, and  
\begin{equation}
    \vec{A'}_1 \cdot d \vec{r}_1 =\left( -{1 \over 4} (i \gamma_2 \gamma_1 + i\bar{\gamma}_2 \bar{\gamma}_1 ) + {1 \over 4} \right) {1 \over 2i} \left( \frac{dz_1}{z_1- z_2 }  - \frac{dz_1^*}{z_1^*- z_2^* }\right).\end{equation}
This leads to the braiding operator 
\begin{equation}
    R = (e^{\pi i/8})^2 \exp{-{\pi \over 4} \gamma_1 \gamma_2} \exp{-{\pi \over 4} \bar{\gamma}_1 \bar{\gamma}_2}.
\end{equation}

Since $\alpha = -1 $ for both layers, we note that, for example, $R| 00\rangle = R_1^{\sigma \sigma} R_1^{\sigma \sigma} | 00 \rangle =  e^{- \pi i/4}$. Repeating this for the other three basis states,  we obtain the full braiding matrix in the $\{ |00 \rangle, |01 \rangle , |10 \rangle , |11 \rangle  \}$ basis:
\begin{equation}
    R = \begin{pmatrix}e^{-i \pi/ 4} & 0 & 0& 0\\
0& e^{i \pi/4} & 0& 0 \\
0 & 0& e^{i \pi /4} & 0 \\
0 & 0& 0& e^{i3 \pi/4}\end{pmatrix}.
\label{braidingmatrix1}
\end{equation}

\begin{figure}[ht]
    \centering
    
    \includegraphics[width = 6cm ]{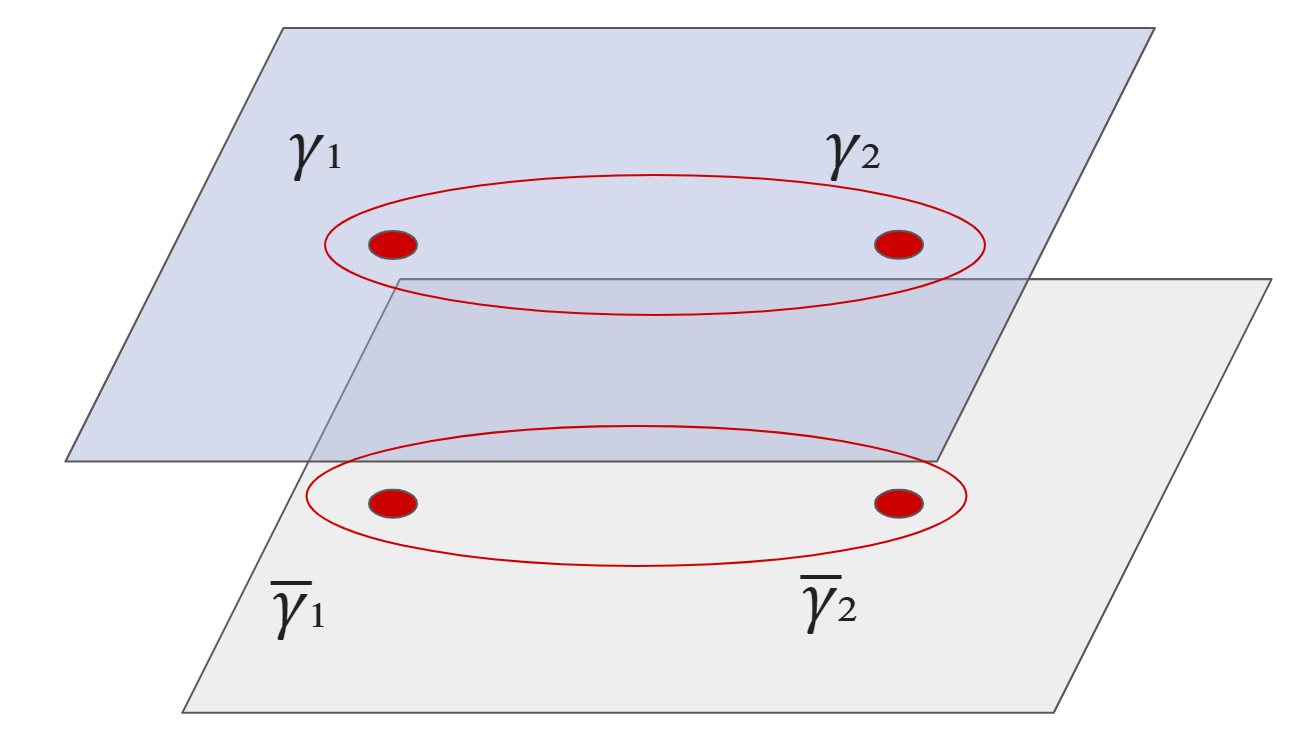}
    \caption{Majorana modes on each layer paired up}
    \label{fig:vortices12}
\end{figure}

After confinement, we are left with the $(\sigma, \sigma)$ anyon, and since it fuses with $(\psi, \psi)$ to itself there is no further identification of anyons needed. The remaining question is the splitting of $(\sigma, \sigma)$ into $a + \bar{a}$.

\begin{figure}[ht]
    \centering
    
    \includegraphics[width = 6cm ]{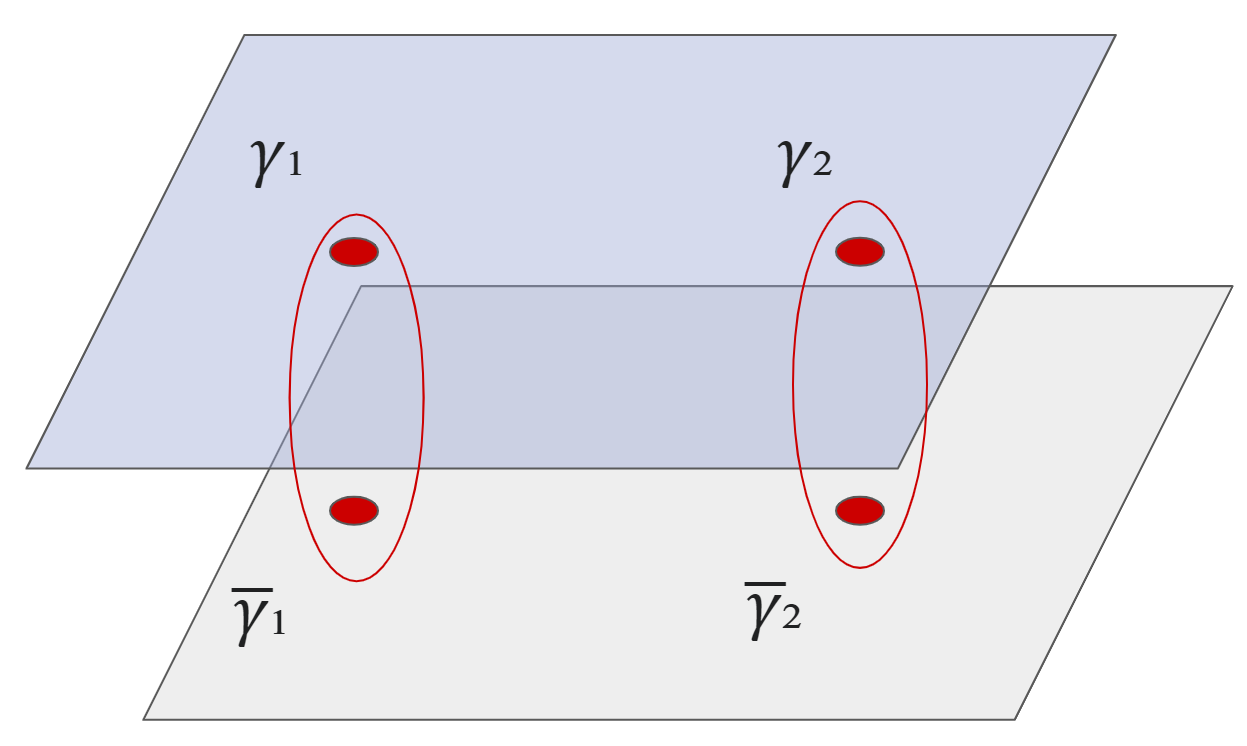}
    \caption{Majorana modes on each vortex paired up}
    \label{fig:vortices11}
\end{figure}

Just as two $\sigma$ anyons on a single layer behave as either $1$ or $\psi$ when zoomed out and considered together (as in Figure \ref{fig:vortices12}), the two $\sigma$ anyons on two different layers behave as either $a$ or $\bar{a}$ when considered together; see Figure \ref{fig:vortices11}. Since they differ by a fermion ($\bar{a} = a \times \psi$ and $a = \bar{a} \times \psi$), we look for eigenstates of the fermionic parity operator localized to a single vortex -- that is, we change the basis from the eigenbasis of $i \gamma_2 \gamma_1$ and $i \bar{\gamma}_2 \bar{\gamma}_1$ to the eigenbasis of $i \bar{\gamma}_1 \gamma_1$ and $i \bar{\gamma}_2 \gamma_2$. With respect to localized fermionic parity, we will denote the even state by  $|a \rangle$ and the odd state by $|\bar{a} \rangle$. The expression for the new basis states in terms of the old basis is given by:
\begin{align}
|aa \rangle ={|01 \rangle - i |10 \rangle \over \sqrt{2}} \nonumber \\
|a \bar{a} \rangle = { |00 \rangle - i |11 \rangle \over \sqrt{2} } \nonumber \\
|\bar{a} a \rangle = {|00 \rangle + i |11 \rangle \over \sqrt{2}} \nonumber \\
|\bar{a} \bar{a} \rangle = {|01 \rangle + i |10 \rangle \over \sqrt{2} }.
\label{basischange}
\end{align}

Under braiding, $|aa \rangle$ and $|\bar{a} \bar{a} \rangle $ transform with a phase of $e^{\pi i/4}$ while 
\begin{align}
    |a \bar{a} \rangle \mapsto e^{-i \pi  /4} |\bar{a} a \rangle \nonumber \\ 
    | \bar{a} a \rangle \mapsto e^{-  i \pi  /4}|a \bar{a} \rangle. 
    \label{aatransform}
\end{align} 
These are the correct braiding coefficients for $\cP_2$, Eq. \eqref{Z4braiding}. 

Note that $|aa \rangle$ should belong to the superselection sector $\psi$ since $a \times a = \psi$, and since $|aa \rangle$ is a linear combination of $|01\rangle$ and $|10\rangle$, each of which belongs to the superselection sector $\psi$ from $1 \times \psi =  \psi$, this is consistent with the fusion rules Eq. \eqref{Z4fusion}. The same holds for the other three states, and this confirms that the $\nu =2 $ phase indeed has $\ZZ_4$ fusion rules, as opposed to $\ZZ_2 \times \ZZ_2$ fusion rules.\footnote{Another way to pin down the fusion rules is the following: we can distinguish between the two fusion rules by noting that two vortices of the same type fuse to $1$ in the $\ZZ_2 \times \ZZ_2$ case but they fuse to $\psi$ in the $\ZZ_4$ case; $1$ will of course have trivial braiding, while $\psi$ will acquire a phase $-1$ under exchange. In the former case, we need $R_1^{ee} = 1^{1/4}$, and in the latter case, $R_{\psi}^{aa} = (-1)^{1/4}.$  In the $\nu = 2$ phase, $|aa \rangle \mapsto e^{i \pi /4} |aa \rangle$ under braiding, and since $(e^{\pi i/ 4})^4 = -1$, it is of $\ZZ_4$ type.}

\subsubsection[411]{Effective Hamiltonian for $\nu = 2$}
\label{Ham2}

Let us discuss what the effective Hamiltonian looks like. The stacked Hamiltonian acts on $\CC^4$. If we take the $|aa \rangle$ sector, 
\begin{equation}
    \left ( -{1 \over 4} (i \gamma_2 \gamma_1 +  i\bar{\gamma}_2 \bar{\gamma}_1)   + {1 \over 4}\right)| aa \rangle = {1 \over 4} \left( |01 \rangle - |01 \rangle -i ( - |10 \rangle   + | 10 \rangle  )\right) + {1 \over 4} |aa\rangle = {1 \over 4} |aa \rangle.
\end{equation} Hence, in this sector, we get the Hamiltonian
\begin{align}
    H = {1 \over 2m} \left( (\vec{p}_1 - \vec{A}_1)^2  + (\vec{p}_2 - \vec{A}_2)^2\right) + (\mathrm{potential}) \nonumber \\
    \vec{A'}_1 \cdot d \vec{r}_1 ={1 \over 4} \frac{1}{2i} \left( \frac{dz_1}{z_1- z_2 }  - \frac{dz_1^*}{z_1^*- z_2^* }\right).
    \label{aaHamiltonian}
 \end{align}

This Hamiltonian clearly reproduces the phase $e^{i \pi /4}$ under braiding as it should. The $|\bar{a} \bar{a} \rangle$ sector works similarly, and the end result is the same. On the other hand, when we consider $| a \bar{a} \rangle$, we need to consider it together with  $| \bar{a} a \rangle$ since $|a \bar{a} \rangle$ and $|\bar{a} a \rangle$ transform into each other after braiding.  The sector spanned by $|a \bar{a} \rangle$ and $|\bar{a} a \rangle$ is $\CC^2 = \mathrm{Span} \{ |00 \rangle, |11 \rangle \}$, and restricting to this to the subspace we see that the gauge field takes the form  
\begin{equation}
     \vec{A'}_1 \cdot d \vec{r}_1 ={1 \over 4} \begin{pmatrix} - 1 & 0 \\ 0 & 3 \end{pmatrix} \frac{1}{2i} \left( \frac{dz_1}{z_1- z_2 }  - \frac{dz_1^*}{z_1^*- z_2^* }\right).\end{equation}
in the $|00\rangle$ and $|11 \rangle$ basis.
 
Now we convert this to the $|a\bar{a} \rangle$ and $|\bar{a} a \rangle$ basis. Noting how $\left ( -{1 \over 4} (i \gamma_2 \gamma_1 +  i\bar{\gamma}_2 \bar{\gamma}_1)   + {1 \over 4}\right) $ acts on the two states, we see that in this basis the gauge field takes the form
\begin{equation}
    \vec{A'}_1 \cdot d \vec{r}_1 ={1 \over 4} \begin{pmatrix} 1 & -2  \\ -2 & 1 \end{pmatrix} \frac{1}{2i} \left( \frac{dz_1}{z_1- z_2 }  - \frac{dz_1^*}{z_1^*- z_2^* }\right). 
\end{equation}

The Hamiltonian describing the interaction between an $a$ vortex and a $\bar{a}$ vortex will take the form 
\begin{equation}
    H = {1 \over 2m} \left( (\vec{p}_1 - \vec{A}_1)^2  + (\vec{p}_2 - \vec{A}_2)^2\right) + (\mathrm{potential})
 \end{equation}
in the $|a\bar{a}\rangle, |\bar{a} a \rangle$ basis, where $\vec{A}$ takes the above form. 

The Berry phase from the gauge field for a half braid results in 
\begin{equation}
    e^{i \pi [{1 \over 4 } (\mathds{1} - 2 \sigma^x)]} = e^{i \pi /4} e^{-i \pi /2} \sigma^x 
\end{equation}
acting on  the states $|a \bar{a} \rangle$ and $|\bar{a} a \rangle$, and this reproduces Eq. \eqref{aatransform}. 
 
\subsection[42]{Stacking $\nu = 1$ with $\nu = -1$}

Now we consider stacking $\cP_1 $ with $\cP_{-1}$. The general structure of the argument is the same as for the stacking of $\cP_1$ with itself, but there are slight differences which will lead to a $\ZZ_2 \times \ZZ_2$ theory (representative of cases where we end up with $\nu = 0 $ mod $4$). The particular case of $\nu = 0$ also has time-reversal symmetry and is related to Class DIII systems, which will be discussed in Sec. \ref{sec:TR}. 

We start with $H$ for two vortices in the phase $\cP_1$:
\begin{equation}
    H = {1 \over 2m} \left( (\vec{p}_1 - \vec{A}_1)^2  + (\vec{p}_2 - \vec{A}_2)^2\right) + (\mathrm{potential})
\end{equation}
and add a second layer in the phase $\cP_{-1}$:
\begin{equation}
    \tilde{H} = {1 \over 2m} \left( (\vec{\tilde{p}}_1 - \vec{\tilde{A}}_1)^2  + (\vec{\tilde{p}}_2 - \vec{\tilde{A}}_2)^2\right)   +(\mathrm{potential}) 
\end{equation}
where we denote by $\tilde{\gamma}_1$ and $\tilde{\gamma}_2$ the Majorana modes of the second layer. 

After stacking, the total Hamiltonian is again $H' = H \otimes \mathds{1} + \mathds{1} \otimes \tilde{H}$, acting on $\left( L^2(\RR^2) \right)^{\otimes 4}\otimes \CC^4$. The condensation process proceeds in the same way as in the $\cP_1 \boxtimes_f \cP_1$ case, and after confinement we obtain:
\begin{equation}
    H' = {1 \over 2m} \left( (\vec{p'}_1 - \vec{A'}_1)^2  + (\vec{p'}_2 - \vec{A'}_2)^2\right) + (\mathrm{potential})
    \end{equation}
where $\vec{A'}_1 \cdot d \vec{r}_1 ={1 \over 4} \frac{(\gamma_1 \gamma_2 + \tilde{\gamma}_1 \tilde{\gamma}_2 )}{2i} \left( \frac{dz_1}{z_1- z_2 }  - \frac{dz_1^*}{z_1^*- z_2^* }\right)$. Note that the overall phase factors $\theta(\nu)$ cancel each other out.

The braiding matrix will then be 
\begin{equation}
R =     \exp{-{\pi \over 4} \gamma_1 \gamma_2} \exp{-{\pi \over 4} \tilde{\gamma}_1 \tilde{\gamma}_2 }.
   \end{equation}

We follow the same steps as in the $\cP_1 \boxtimes_f \cP_1$ case. Now, $\alpha = -1 $ for the first layer and $\alpha = +1$ for the second layer, so the state $|0 0\rangle = |0 \rangle \otimes |0 \rangle$ corresponds to the state in the fusion channel $1$ on the first layer and $\psi$ on the second layer. Thus we have, for example, $R | 00 \rangle = (R^{\nu = 1})_{1}^{\sigma \sigma} (R^{\nu = -1})_{\psi}^{\sigma \sigma} | 00 \rangle =   e^{-\pi i/2} |00 \rangle$. Repeating this for the other states, we compute the braiding matrix in this basis to be
\begin{equation}
    R = \begin{pmatrix} - i & 0 & 0 & 0 \\
0 & 1 & 0 & 0 \\
0 & 0 & 1 & 0 \\
0 & 0 & 0 & i\end{pmatrix}. \end{equation}

Now we consider the system in a different basis: instead of thinking of $\CC^4$ as $\CC^2_{\text{layer} 1} \otimes \CC^2_{\text{layer} 2} $, we think of it as $\CC^2_{\text{vortex} 1} \otimes \CC^2_{\text{vortex} 2}$, where each vortex carries two Majorana modes $\gamma_i$, $\tilde{\gamma}_i.$ Each vortex carries a space $\CC^2$ whose states are eigenstates of the vortex-localized fermionic parity operator $i  \tilde{\gamma}_1 \gamma_1$ or $i  \tilde{\gamma}_2 \gamma_2$. On each $\CC^2_{\text{vortex} i}$ we have an even state $|e \rangle $ and an odd state $|m \rangle$; the total fermionic Hilbert space $\CC^4$ is spanned by the basis $|ee \rangle \equiv |e \rangle_1 \otimes |e \rangle_2$, $|em \rangle$, $|me \rangle$, and $|mm \rangle$. We can write these states in terms of the old basis states as:
\begin{align}
|ee \rangle &={|01 \rangle - i |10 \rangle \over \sqrt{2}}  \nonumber \\
|em \rangle &= { |00 \rangle - i |11 \rangle \over \sqrt{2} } \nonumber \\
|me \rangle &= {|00 \rangle + i |11 \rangle \over \sqrt{2}}  \nonumber \\
|mm \rangle &= {|01 \rangle + i |10 \rangle \over \sqrt{2} }.
\label{basischange2}
\end{align}

Since we know how the states $|00 \rangle$, $|01 \rangle$, etc. transform under braiding, we can compute the behavior of the new basis states under braiding. We see that $R_1^{ee} = R_1^{mm} = 1$; and also that $R |em \rangle = -i |m e \rangle$ and $R |me \rangle = -i | em \rangle$, from which we see that $R_1^{em} R_1^{me} = M_1^{em} = -1$. These are indeed the correct braiding coefficients for the toric code, Eq. \eqref{Z2Z2braiding}.  

We also note that $|01  \rangle$, for example, corresponds to fusion channel $1$ on both layers (this is different from the $\cP_1 \boxtimes_f \cP_1$ case, since $\alpha$ is now different for each layer). Hence $e \times e = 1$, which is consistent with $\ZZ_2 \times \ZZ_2$ rather than $\ZZ_4$ fusion rules.\footnote{Again, we can also confirm this by looking at the braiding phases. Since the $\nu = 0 $ phase has $|ee \rangle \mapsto |ee \rangle$ under braiding, and $(R^{vv})^4 = 1^4 = 1$ so it is indeed of $\ZZ_2 \times \ZZ_2 $ type.} 

The effective Hamiltonians involving different types of vortices can be obtained from this braiding matrix in the same manner as the $\cP_1 \boxtimes_f \cP_1$ case.

\subsection{Action of time-reversal and Class DIII superconductors}
\label{sec:TR}

In the $\cP_1 \boxtimes_f \cP_{-1}$ system, we have time-reversal (TR) symmetry which acts as \cite{Bernevig}
\begin{align}
    \gamma_i \mapsto - \tilde{\gamma}_i \nonumber \\
    \tilde{\gamma}_i \mapsto \gamma_i.
\end{align}
This flips the sign of the fermionic parity operator on each vortex:
\begin{equation}
i \tilde{\gamma}_i \gamma_i \mapsto -i \tilde{\gamma}_i \gamma_i
\end{equation} 
Hence, $|e \rangle$ and $|m \rangle$ map to each other under time-reversal.

Stacking a $p+ip $ (which belongs to $\cP_1$) and a $p-ip$ superconductor (which belongs to $\cP_{-1}$), we obtain a superconductor in Class DIII, a system that is protected by time-reversal symmetry from deformation to the trivial system. If we break time-reversal symmetry, we can deform it to the $s$-wave superconductor, which has the toric code as its underlying topological order \cite{Hansson2004}. In the $s$-wave superconductor, the vortex $m$ and the sector which has a vortex and a fermion $ e = m \times \psi$ are unrelated by time-reversal symmetry, whereas we have seen that in the nontrivial Class DIII TR-invariant superconductor the TR operation exchanges $e$ and $m$. Thus, on the level of the TQFT, this nontrivial TR action distinguishes it from the trivial phase.

Now let us see what happens when we stack the two nontrivial TR-invariant superconductors. Since the underlying topological order is the toric code, we stack two copies and condense the $(\psi, \psi)$ particle. Most combinations are confined -- we are left with 
\begin{align}
    1' = (1, 1) \sim (\psi, \psi) \nonumber \\
    \psi' = (1, \psi) \sim (\psi, 1)  \nonumber \\
    e' = (e, e) \sim (m,m) \nonumber \\
    m' = (e,m ) \sim (m,e)
\end{align}
and the new theory obeys the toric code braiding and fusion rules, as it should. The only difference is in the action of TR: since TR exchanges $e$ and $m$ in the original systems being stacked, we see that the new $e'$ and $m'$ anyons are invariant under TR. Thus we have obtained the the trivial phase by stacking two copies of the nontrivial phase, and this recovers the well-known $\ZZ_2$ classification of Class DIII systems in $2+1$d \cite{ Bernevig, Ryuetal}.

Note how this works from the perspective of effective Hamiltonians: With the two systems stacked, we could have terms like $i \gamma_1 \tilde{\gamma}_1 V_1$ which is now local (unlike $i \gamma_2 \gamma_1 V$). This would break the  degeneracy between the $e$ and $m$ particles, since $i \gamma_1  \tilde{\gamma}_1| e \rangle = + |e \rangle$, $i \gamma_1  \tilde{\gamma}_1| m \rangle = - |m \rangle$. However, under time-reverasl, $i \gamma \tilde{\gamma} \mapsto - i \tilde{\gamma} (-) \gamma = -i \gamma \tilde{\gamma}$, so such terms are not TR-invariant. 

On the other hand if we take e.g. the $\nu = 2$ phase, there is no TR symmetry, so nothing prevents us from adding such terms which would lift the degeneracy between $a$ and $\bar{a}$. As discussed in \cite{BernevigNeupert}, there are no stable Majorana bound states in even $\nu$ phases, unless we protect them by a symmetry.

\section{Stacking: odd from even-odd}

\subsection[51]{$\cP_3= \cP_2 \boxtimes_f \cP_1$}

Let us first consider stacking $\cP_2$ with $\cP_1$. We are stacking the anyons $1, a , \bar{a}, \psi$ of $\cP_2$ with the anyons $1, \sigma, \psi$ of $\cP_1$ and condensing the $(\psi, \psi)$ anyon. Most of the combinations are confined, and we are left with \begin{align}
    1' = (1, 1) \nonumber \\
    \sigma' = (a, \sigma) \sim (\bar{a}, \sigma) \nonumber \\ \psi' = (1, \psi) \sim (\psi, 1)
\end{align}
with the usual fusion rules for the Ising TQFT, Eq. \eqref{Isingfusion}.

As we saw in section \ref{nu11}, there are four different $\nu = 2 $ Hamiltonians for two vortices, corresponding to the sectors $|aa \rangle$, $|a \bar{a} \rangle$, $|\bar{a} a \rangle$, and $|\bar{a} \bar{a} \rangle$; each Hamiltonian acts on a Hilbert space $ L^2(\RR^2) \otimes L^2(\RR^2)$. Consider the $|aa \rangle$ sector, which has the Hamiltonian as in Eq. \eqref{aaHamiltonian}, with gauge field
\begin{equation}
\vec{A'}_1 \cdot d \vec{r}_1 ={1 \over 4} \frac{1}{2i} \left( \frac{dz_1}{z_1- z_2 }  - \frac{dz_1^*}{z_1^*- z_2^* }\right).
\end{equation}

Let us stack this with a $\nu = 1$ system, which is just the Hamiltonian in Eq. \eqref{EH1},
\begin{equation}
    H = {1 \over 2m} \left( (\vec{p}_1 - \vec{A}_1)^2  + (\vec{p}_2 - \vec{A}_2)^2\right) + i\gamma_2 \gamma_1 V(|\vec{r}_1 - \vec{r}_{2}| ) 
   \label{}
\end{equation}
with 
\begin{equation}
     \vec{A}_1 \cdot d\vec{r}_1 = \left(-{i \over 4} \gamma_2 \gamma_1 + {1 \over 8} \right) \left( {1 \over 2i} \frac{dz_1}{z_1 - z_{2}} - {1 \over 2i} \frac{d\bar{z}_1 }{\bar{z}_1 - \bar{z}_2}\right).
\end{equation}
The total Hilbert space becomes $ \left( L^2(\RR^2) \right)^{\otimes 4} \otimes \CC^2$, but after condensation, forcing the vortex ($a$ or $\bar{a}$ on the first layer and $\sigma$ on the second) position to be the same on the two layers, we are left with $\left( L^2(\RR^2) \right)^{\otimes 2} \otimes \CC^2$. The resulting Hamiltonian again takes the general form of Eq. \eqref{EH1}, though $F(\nu)$ and hence the braiding coefficients are now different. This Hamiltonian now describes the interaction of two $(a, \sigma)$ particles, and having an internal fermionic Hilbert space $\CC^2$ is consistent with the fusion rules 

\begin{equation}
    (a, \sigma) \times (a, \sigma) = (\psi, 1) + (\psi, \psi) \sim \psi + 1. 
\end{equation}

 The result actually should be the same if we had started with the $|\bar{a} \bar{a} \rangle$ sector or the sector containing $| \bar{a} a \rangle$ and $|a \bar{a} \rangle$, since $(a, \sigma) \sim (\bar{a}, \sigma).$ Regardless of which Hamiltonian we chose for the $\nu = 2$ phase, after fermionic stacking, we end up with a single type of vortex, described by a Hamiltonian of the type Eq. \eqref{EH1}.

Let us confirm that we get the correct braiding coefficients. First, consider the case where we have started with the $|aa \rangle $ sector. The braiding matrix for the $\nu =1$ phase is $diag(e^{- i \pi /8}, e^{3 \pi i/8}) $ for the braiding of two $\sigma$ vortices. However, for the $\nu =3$ phase we need to switch the two components: the $\nu =3$ vortex $\sigma' = (a, \sigma)$ has fusion 
\begin{equation}
\sigma' \times \sigma' = (\psi, 1) + (\psi, \psi) = \psi' + 1'. \end{equation}
Hence, if we are in the $1$ sector of the $\nu = 1$ phase that is being stacked, we are in the $\psi'$ sector of the $\nu =3$ phase, and vice versa. Thus,  $R^{\sigma' \sigma'}_{1'} = e^{3 \pi i /8}$ and $R^{\sigma' \sigma'}_{\psi'} = e^{-\pi i/8}$ up to the additional phase coming from the $a$s. After multiplying by a phase $e^{i \pi/4}$ from the exchange of two $a$s, we get \begin{align} R^{\sigma' \sigma'}_{1'} = e^{5\pi i/8 } \nonumber \\
 R^{\sigma' \sigma'}_{\psi'} = e^{ \pi i/8}.
 \label{3braid}
 \end{align} 
 These are indeed the braiding coefficients for the $\nu =3$ phase, Eq. \eqref{isingbraid}. Since $R_{\psi}^{\bar{a} \bar{a}} = R_{\psi}^{aa}$, the same argument would hold had we started out in the $|\bar{a} \bar{a} \rangle$ sector.\footnote{If we had worked in the sector consisting of $|a \bar{a} \rangle$ and $|\bar{a} a \rangle$, we may not immediately get the correct coefficients for $R^{\sigma' \sigma'}_{i}$ since $R^{a\bar{a}}_1$ and $R^{\bar{a} a}_1$ do not have  invariant meanings. On the other hand, if we first compute the topological spin of $\sigma'$ and the double-braiding/monodromy coefficients $M^{\sigma' \sigma'}_i$  , which have invariant meanings, and then compute $R^{\sigma' \sigma'}_{i}$, we will arrive at the correct result.}

\subsection{General braiding coefficients from stacking }

Recall that for any odd $\nu$, we have $R_1^{\sigma \sigma} = \theta(\nu) e^{ \alpha \pi i /4}$ and $R_{\psi}^{\sigma \sigma} = \theta(\nu) e^{- \alpha \pi i /4} $, where $\theta (\nu) = e^{\nu \pi i\over 8}$ and $\alpha = -1$ for $\nu = 1$ mod $4$ and $+1$ for $\nu = 3 $ mod  $4$.

The value of $\alpha$ can be understood from the stacking perspective in the following way. A $\nu = 1$ mod $4$ phase is obtained by stacking $\cP_1$  with a $\cP_{4n}$; the latter phase has $e$ and $m$ type vortices. After stacking, we get the vortex $\sigma' = (\sigma, e) \sim (\sigma, m)$, with the fusion rule  \begin{equation}
\sigma' \times \sigma' = (1, 1) + (\psi, 1) = 1' + \psi'\end{equation}
so the sectors $1'$ and $\psi'$ of $\cP_{4n+1}$ correspond to the sectors $1$ and $\psi$ of  $\cP_1$. Hence we get $\alpha = -1$ (since $\cP_1$ has $\alpha = -1$).

On the other hand, $\cP_{4n+3} = \cP_1 \boxtimes_f \cP_{4n+2}$, and since $\cP_{4n+2}$ has $a$ and $\bar{a}$ type vortices, $\cP_{4n+3}$ has the vortex $\sigma' = (\sigma, a) \sim (\sigma, \bar{a})$ with the fusion rule  
\begin{equation}
 \sigma' \times \sigma' = (1, \psi) + (\psi, \psi) = \psi' + 1', 
\end{equation} 
so the sectors $1'$ and $\psi'$ of $\cP_{4n+3} $ correspond to the sectors $\psi$ and $1$ of $\cP_1$ respectively. This means that the braiding coefficients $R_1^{\sigma \sigma}$ and $R_{\psi}^{\sigma \sigma}$ need to change places, compared to those for $\cP_1$ (and $\theta(\nu) $ is unaffected since it is common to both). Thus we see that $\alpha = +1$ for $\cP_{4n+3}.$

We can also think of an odd phase $\cP_{\nu}$ as the stacking of $\cP_2$ with some other odd phase $\cP_{\nu - 2}$. By the above logic, stacking with $\cP_2$ changes the sign of $\alpha$; on the the hand, the braiding coefficients for the vortex $\sigma'$ of $\cP_{\nu}$ also acquires a phase $e^{i \pi /4}$ from the braiding of the $a$ vortices of $\cP_2$. Hence the overall phase behaves as
\begin{equation}
\theta(\nu) = e^{2\pi i/8} \theta(\nu - 2).
\end{equation}

Thus we see that, whenever $\nu$ advances by $2$, going from an odd phase to an odd phase, the value of $\alpha$ gets reversed and $\theta(\nu)$ increases by $e^{2 \pi i/8} $. This means that once we are given the braiding coefficients for one odd phase, we can obtain those of all the other odd phases immediately.

\section*{Acknowledgements}

I am grateful to Anton Kapustin for discussions and comments. I would also like to thank Yu-An Chen and Po-Shen Hsin for helpful discussions. The work is supported
by the U.S. Department of Energy, Office of Science, Office of High Energy Physics, under Award Number de-sc0011632.

\printbibliography

@article{BGK,
   title={State sum constructions of spin-TFTs and string net constructions of fermionic phases of matter},
   volume={2017},
   ISSN={1029-8479},
   url={http://dx.doi.org/10.1007/JHEP04(2017)096},
   DOI={10.1007/jhep04(2017)096},
   number={4},
   journal={Journal of High Energy Physics},
   publisher={Springer Science and Business Media LLC},
   author={Bhardwaj, Lakshya and Gaiotto, Davide and Kapustin, Anton},
   year={2017},
   month={Apr}
}

@article{Ryuetal,
   title={Classification of topological quantum matter with symmetries},
   volume={88},
   ISSN={1539-0756},
   url={http://dx.doi.org/10.1103/RevModPhys.88.035005},
   DOI={10.1103/revmodphys.88.035005},
   number={3},
   journal={Reviews of Modern Physics},
   publisher={American Physical Society (APS)},
   author={Chiu, Ching-Kai and Teo, Jeffrey C.Y. and Schnyder, Andreas P. and Ryu, Shinsei},
   year={2016},
   month={Aug}
}

@book{Bernevig, author = "Bernevig, B.A. ", title = "Topological Insulators and Superconductors",  publisher = {Princeton University Press},year = 2013}

@article{ReadGreen,
   title={Paired states of fermions in two dimensions with breaking of parity and time-reversal symmetries and the fractional quantum Hall effect},
   volume={61},
   ISSN={1095-3795},
   url={http://dx.doi.org/10.1103/PhysRevB.61.10267},
   DOI={10.1103/physrevb.61.10267},
   number={15},
   journal={Physical Review B},
   publisher={American Physical Society (APS)},
   author={Read, N. and Green, Dmitry},
   year={2000},
   month={Apr},
   pages={10267–10297}
}

@article{Ivanov,
   title={Non-Abelian Statistics of Half-Quantum Vortices inp-Wave Superconductors},
   volume={86},
   ISSN={1079-7114},
   url={http://dx.doi.org/10.1103/PhysRevLett.86.268},
   DOI={10.1103/physrevlett.86.268},
   number={2},
   journal={Physical Review Letters},
   publisher={American Physical Society (APS)},
   author={Ivanov, D. A.},
   year={2001},
   month={Jan},
   pages={268–271}
}

@article{KitaevAnyons,
   title={Anyons in an exactly solved model and beyond},
   volume={321},
   ISSN={0003-4916},
   url={http://dx.doi.org/10.1016/j.aop.2005.10.005},
   DOI={10.1016/j.aop.2005.10.005},
   number={1},
   journal={Annals of Physics},
   publisher={Elsevier BV},
   author={Kitaev, Alexei},
   year={2006},
   month={Jan},
   pages={2–111}
}

@article{Burnell,
   title={Anyon Condensation and Its Applications},
   volume={9},
   ISSN={1947-5462},
   url={http://dx.doi.org/10.1146/annurev-conmatphys-033117-054154},
   DOI={10.1146/annurev-conmatphys-033117-054154},
   number={1},
   journal={Annual Review of Condensed Matter Physics},
   publisher={Annual Reviews},
   author={Burnell, F.J.},
   year={2018},
   month={Mar},
   pages={307–327}
}

@article{Hansson2004,
   title={Superconductors are topologically ordered},
   volume={313},
   ISSN={0003-4916},
   url={http://dx.doi.org/10.1016/j.aop.2004.05.006},
   DOI={10.1016/j.aop.2004.05.006},
   number={2},
   journal={Annals of Physics},
   publisher={Elsevier BV},
   author={Hansson, T.H. and Oganesyan, Vadim and Sondhi, S.L.},
   year={2004},
   month={Oct},
   pages={497–538}
}

@misc{BernevigNeupert,
    title={Topological Superconductors and Category Theory},
    author={Bernevig, A. and Neupert, T.},
    year={2015},
    eprint={1506.05805},
    archivePrefix={arXiv},
    primaryClass={cond-mat.str-el}
}

@article{LanKongWen1,
   title={Theory of (2+1)-dimensional fermionic topological orders and fermionic/bosonic topological orders with symmetries},
   volume={94},
   ISSN={2469-9969},
   url={http://dx.doi.org/10.1103/PhysRevB.94.155113},
   DOI={10.1103/physrevb.94.155113},
   number={15},
   journal={Physical Review B},
   publisher={American Physical Society (APS)},
   author={Lan, Tian and Kong, Liang and Wen, Xiao-Gang},
   year={2016},
   month={Oct}
}

@article{LanKongWen2,
   title={Modular Extensions of Unitary Braided Fusion Categories and 2+1D Topological/SPT Orders with Symmetries},
   volume={351},
   ISSN={1432-0916},
   url={http://dx.doi.org/10.1007/s00220-016-2748-y},
   DOI={10.1007/s00220-016-2748-y},
   number={2},
   journal={Communications in Mathematical Physics},
   publisher={Springer Science and Business Media LLC},
   author={Lan, Tian and Kong, Liang and Wen, Xiao-Gang},
   year={2016},
   month={Sep},
   pages={709–739}
}

@article{bosoncondensation,
   title={Boson condensation in topologically ordered quantum liquids},
   volume={93},
   ISSN={2469-9969},
   url={http://dx.doi.org/10.1103/PhysRevB.93.115103},
   DOI={10.1103/physrevb.93.115103},
   number={11},
   journal={Physical Review B},
   publisher={American Physical Society (APS)},
   author={Neupert, Titus and He, Huan and von Keyserlingk, Curt and Sierra, Germán and Bernevig, B. Andrei},
   year={2016},
   month={Mar}
}

@article{splitting,
   title={Splitting of Majorana-Fermion Modes due to Intervortex Tunneling in apx+ipySuperconductor},
   volume={103},
   ISSN={1079-7114},
   url={http://dx.doi.org/10.1103/PhysRevLett.103.107001},
   DOI={10.1103/physrevlett.103.107001},
   number={10},
   journal={Physical Review Letters},
   publisher={American Physical Society (APS)},
   author={Cheng, Meng and Lutchyn, Roman M. and Galitski, Victor and Das Sarma, S.},
   year={2009},
   month={Aug}
}

@misc{Kitaevtalk2, 
author = {Kitaev, A.},
title = {Toward Topological Classification of Phases with Short-range Entanglement},
url = {http://online.kitp.ucsb.edu/online/topomat11/kitaev/rm/jwvideo.html},
year = {2011}
}

@article{BaisSlingerland,
   title={Condensate-induced transitions between topologically ordered phases},
   volume={79},
   ISSN={1550-235X},
   url={http://dx.doi.org/10.1103/PhysRevB.79.045316},
   DOI={10.1103/physrevb.79.045316},
   number={4},
   journal={Physical Review B},
   publisher={American Physical Society (APS)},
   author={Bais, F. A. and Slingerland, J. K.},
   year={2009},
   month={Jan}
}

@article{Laughlin2,
  title = {Anomalous Quantum Hall Effect: An Incompressible Quantum Fluid with Fractionally Charged Excitations},
  author = {Laughlin, R. B.},
  journal = {Phys. Rev. Lett.},
  volume = {50},
  issue = {18},
  pages = {1395--1398},
  numpages = {0},
  year = {1983},
  month = {May},
  publisher = {American Physical Society},
  doi = {10.1103/PhysRevLett.50.1395},
  url = {https://link.aps.org/doi/10.1103/PhysRevLett.50.1395}
}

@article{MooreRead,
title = "Nonabelions in the fractional quantum hall effect",
journal = "Nuclear Physics B",
volume = "360",
number = "2",
pages = "362 - 396",
year = "1991",
issn = "0550-3213",
doi = "https://doi.org/10.1016/0550-3213(91)90407-O",
url = "http://www.sciencedirect.com/science/article/pii/055032139190407O",
author = "Gregory Moore and Nicholas Read"
}

@article{OneformGauging,
   title={Comments on one-form global symmetries and their gauging in 3d and 4d},
   volume={6},
   ISSN={2542-4653},
   url={http://dx.doi.org/10.21468/SciPostPhys.6.3.039},
   DOI={10.21468/scipostphys.6.3.039},
   number={3},
   journal={SciPost Physics},
   publisher={Stichting SciPost},
   author={Hsin, Po-Shen and Lam, Ho Tat and Seiberg, Nathan},
   year={2019},
   month={Mar}
}

\end{document}